\documentclass{article}
\usepackage{spconf,amsmath,graphicx}
\usepackage{epstopdf}
\usepackage{amssymb}
\usepackage{xspace}
\usepackage{makecell}
\usepackage{array}
\usepackage{booktabs}
\usepackage{url}

\small \title{Deep Segment Attentive embedding for duration robust speaker verification}
\small \name{Bin Liu$^{1,2}$, Shuai Nie$^{1}$,  Yaping Zhang$^{1,2}$, Shan Liang$^1$, Wenju Liu$^1$}
\small \address{$^1$ National Laboratory of Patten Recognition, Institute of Automation, Chinese Academy of Sciences, China\\
  $^2$ School of Artificial Intelligence, University of Chinese Academy of Sciences, China\\
  \small\texttt{\{bin.liu2015,shuai.nie,yaping.zhang,sliang,lwj\}@nlpr.ia.ac.cn}}

\begin{document}
%
\maketitle
\begin{abstract} \small
LSTM-based speaker verification usually uses a fixed-length local segment randomly truncated from an utterance to learn the utterance-level speaker embedding, while using the average embedding of all segments of a test utterance to verify the speaker, which results in a critical mismatch between testing and training. This mismatch degrades the performance of speaker verification, especially when the durations of training and testing utterances are very different. To alleviate this issue, we propose the deep segment attentive embedding method to learn the unified speaker embeddings for utterances of variable duration. Each utterance is segmented by a sliding window and LSTM is used to extract the  embedding of each segment. Instead of only using one local segment, we use the whole utterance to learn the utterance-level embedding by applying an attentive pooling to the embeddings of all segments. Moreover, the similarity loss of segment-level embeddings is introduced to guide the segment attention to focus on the segments with more speaker discriminations, and jointly optimized with the similarity loss of utterance-level embeddings. Systematic experiments on Tongdun and VoxCeleb show that the proposed method significantly improves robustness of duration variant and achieves the relative Equal Error Rate reduction of 50\% and 11.54\% , respectively.

\end{abstract}
\begin{keywords} \small
deep segment attentive embedding, speaker verification, duration robustness, LSTM
\end{keywords}
\section{Introduction} \small
\label{sec:intro}
The key to speaker verification is to extract the utterance-level speaker vectors with a fixed dimension for utterances of variable duration. The extracted  speaker vector is expected to be as close as possible to the same speaker while far from other speakers. It remains a challenge to extract the robust speaker vectors for utterances of variable duration, especially when the utterance duration varies greatly. The i-vector/PLDA framework \cite{Dehak2011Front,Prince2007Probabilistic, Cumani2013Probabilistic} can easily extract the fixed dimension speaker vectors for utterances of arbitrary duration using statistical modeling. But it suffers performance reduction when handling short utterances \cite{li2017deep,snyder2017deep}. The reason is that i-vector is a Gaussian-based statistical feature, whose estimation need sufficient samples. And the short utterance will lead to the uncertainty in the estimated i-vector.

Deep learning based speaker embedding \cite{li2017deep,Variani2014Deep,wan2018generalized} is another mainstream approach to speaker verification, which has been extensively studied recently and achieved promising performance in short-duration text-independent task. There are two ways to extract speaker embeddings using deep models. One approach is averaging bottleneck features from frame-level speaker classification networks \cite{Variani2014Deep}. Another approach is directly learning utterance-level speaker embeddings with distance-based similarity loss, such as triplet loss \cite{li2017deep,zhang2017end} and generalized end-to-end (GE2E) loss \cite{wan2018generalized}.

LSTM-based speaker embedding is one of the most important deep speaker verification methods and has been demonstrated to be substantially promising \cite{Sainath2015Convolutional,Heigold2015End}. Owing to the powerful ability in modeling time-series data, LSTM can effectively capture the local correlation information of speech, which is very important for speaker verification. But it is still challenging for LSTM to model the long-term dependency of utterances, especially very long utterances. In addition, in order to facilitate batch training, LSTM-based speaker verification usually uses a fixed-length local segment randomly truncated from an utterance to learn the utterance-level speaker embedding in training phase, while using the average embedding of all segments of a test utterance to verify the speaker in testing phase, which leads to a critical mismatch between testing and training. The mismatch dramatically degrades the performance of speaker verification, especially when the difference of durations between training and testing utterances is large. Many methods are proposed to handle the issue of duration variability. The attention-based pooling \cite{okabe2018attentive,zhu2018self} is one of the most important technologies. But most of the attention mechanisms are performed at the frame level, which will leads to the ``over-average'' problem, especially when the utterance is very long.

To alleviate this issue, we propose the deep segment attentive embedding method to learn the unified speaker embeddings for utterances of variable duration. For both training and testing, we use a sliding window to divide utterances into the fixed-length segments and then use LSTM to extract the embedding of each segment. Finally, all segment-level embeddings of an utterance are pooled into a fixed-dimension vector through the segment attention, which is used as the utterance-level speaker embedding. The similarity loss of utterance-level embeddings is used to train the whole network. In addition, in order to guide the segment attention to focus on the segments with more speaker discriminations, we further incorporate the similarity loss of segment-level embeddings. With the joint optimization of the segment-level and utterance-level similarity loss, both local details and global information of utterances are taken into account. Instead of only using one local segment, we use the whole utterance to learn the utterance-level embedding, which unifies the process of training and testing and avoids the mismatch between them.

\section{Related Work} 
\label{sec:relatedwork}

There are some efforts on the issue of duration variability. For example, in the conventional i-vector systems, \cite{Kenny2013PLDA} proposed to propagate the uncertainty relevant to the i-vector extraction process into the PLDA model, which better handled the duration variability. Moreover, in the deep learning based speaker embedding systems, the complementary center loss is proposed in \cite{Na2018Deep, Nam2018Robust, Sarthak2018Learning} in order to solve the problem of large variation in text-independent utterances, including the duration variation. It acts as a regularizer that reduces the intra-class distance variance of the final embedding vectors. However, they don't explicitly model the duration variability of utterances and the mismatch between training and testing phase still exists.

Furthermore, attention mechanisms  have been utilized to capture the long-term variations of speaker characteristics in \cite{okabe2018attentive,zhu2018self}. An important metric is computed by the attention network, which is used to calculate the weighted mean of the frame-level embedding vectors. However, most of the attention mechanisms are performed at the frame level, which will leads to the ``over-average" problem, especially when the utterance is very long.

\section{Proposed Approach}
\label{sec:Deep-Segment-Attentive-Embeddings}

It is still challenging for LSTM to model the long-term dependency of utterances, especially very long utterances. And the mismatch between training and testing phase degrades the performance of speaker verification, especially when the difference of durations between training and testing utterances is large. Therefore, we propose the deep segment attentive embedding method to extract the unified speaker embeddings for utterances of  variable duration.

\begin{figure}[tb]
  \centering
  \includegraphics[width=0.42\textwidth]{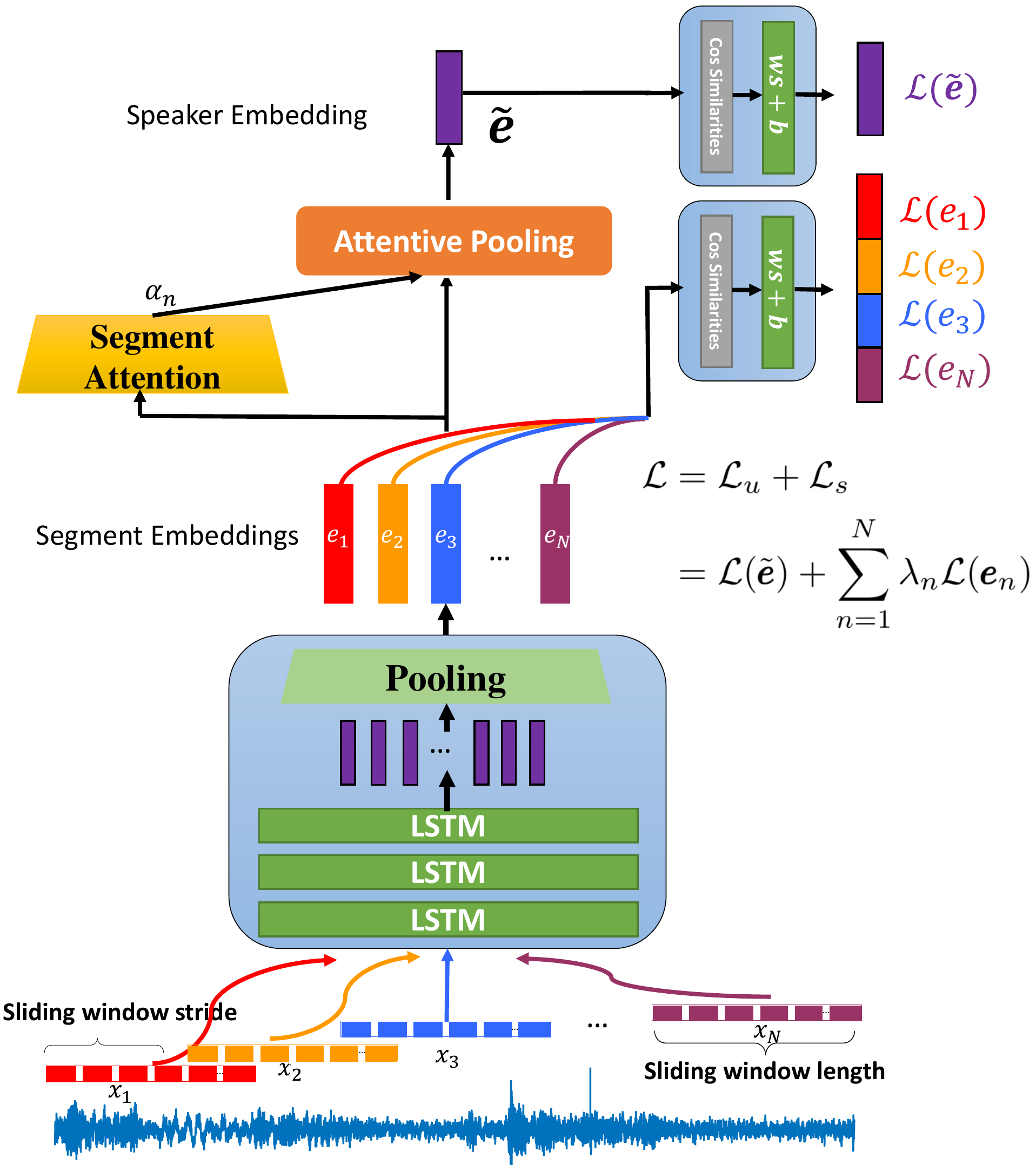}\\
  \caption{\small System overview. For each batch training, there are $Q \times P$ utterances from $Q$ different speakers and each speaker has $P$ utterances. We only draw one utterance for simplicity.}
  \label{fig:System-overview}
\end{figure}

As is shown in Fig.~\ref{fig:System-overview}, we use a sliding window with $50\%$ overlap to divide utterances into the fixed-length segments and LSTM is used to extract the embedding of each segment. Finally, all segment-level embeddings of an utterance are pooled into a fixed-dimension utterance-level speaker embedding through the segment attention mechanism. The whole network is trained with the joint supervision of the utterance-level and segment-level similarity loss. It can extract the unified speaker embeddings for utterances of variable duration and take into account both local details and global information of utterances, especially long utterances.

\subsection{Deep segment attentive embedding}

For both training and testing, we use a sliding window with $50\%$ overlap to divide an utterance into the fixed-length segments. Supposed that we  get $N$ speech segments $\mathbf{X} = \{\pmb{x}_{1}, \pmb{x}_{2}, \cdots , \pmb{x}_{N}\}$. The sliding window length $T$ is randomly chosen within $[80,120]$ frames but the length of segments in a batch is fixed. The vector $\mathbf{x}_{n}^{t}$ represents the feature of segment $n$ at frame $t$, which is fed into the network and the output is $\mathbf{h}_{n}^{t}$. The last frame of output is used as the segment representation $f(\mathbf{x}_{n};\mathbf{w})=\mathbf{h}_{n}^{T}$, where $\mathbf{w}$ represents parameters of the network. The segment-level speaker  embedding is defined as the $L_2$ normalization of the segment representation:
\begin{equation}
\label{equ_eji}
\mathbf{e}_{n} = \frac{f(\mathbf{x}_{n};\mathbf{w}) }{\left \| f(\mathbf{x}_{n};\mathbf{w}) \right \|_{2}}.
\end{equation}
We compute the embedding vector of each segment according to Eq.~\ref{equ_eji} $\mathbf{E} = \{\pmb{e}_{1}, \pmb{e}_{2}, \cdots , \pmb{e}_{N}\}$. Let the dimension of the segment-level speaker embedding $\pmb{e}_n$ be $d_e$.

It is often the case that some segment-level embeddings are more relevant and important for discriminating speakers than others. We therefore apply attention mechanisms to integrate the segment embeddings by automatically calculating the importance of each segment. For each segment-level embedding $\pmb{e}_n$, we could learn a score $\pmb{\alpha}_{n}$ using the segment attention mechanism. All segment-level embeddings of an utterance are pooled into a fixed-dimension utterance-level speaker embedding through the segment attention mechanism.

For each segment embedding $\pmb{e}_{n}$, we apply the multi-head attention mechanism \cite{lin2017structured} to learn a score $\pmb{\alpha}_{n}$ as follows:

\begin{equation}
\label{equ_att_ht}
\pmb{\alpha}_{n} = \text{softmax} \left( g(\pmb{e}_n \mathbf{W}_1)\mathbf{W}_2 \right),
\end{equation}
where $\mathbf{W}_1$ and $\mathbf{W}_2$ are parameters of the multi-head attention mechanism; $\mathbf{W}_1$ is a matrix of size $d_e \times d_a$; $\mathbf{W}_2$ is a matrix of size $d_a \times d_r$; $d_a$ is the attention dim and $d_r$ is a hyperparameter that represents the number of attention heads; $g(\cdot)$ is the ReLU activation function \cite{nair2010rectified}. When the number of attention heads $d_r= 1$, it is simply a basic attention. The normalized weight $\pmb{\alpha}_{n} \in [0,1]$  is computed by the softmax function. The weight vector is then used in the attentive pooling layer to calculate the utterance-level speaker embedding $\tilde{\pmb{e}}$:
\begin{equation}
\label{equ_e}
\tilde{\pmb{e}} = \sum_{n=1}^{N}\pmb{\alpha}_{n} \pmb{e}_n.
\end{equation}

When the number of attention heads $d_r = 1$, $\tilde{\pmb{e}}$ is simply a weighted mean vector computed from $\mathbf{E}$, which is expected to reflect an aspect of speaker discriminations in the given utterance. Obviously, speakers can be discriminated along multiple aspects, especially when the utterance duration is long. By increasing $d_r$, we can easily have multiple attention heads to focus on different pattern aspects from an utterance. In order to encourage diversity in the attention vectors, \cite{zhu2018self} introduced a penalty term $\mathcal L_{p}$ when $d_r > 1$:
\begin{equation}
\label{equ_lp}
\mathcal L_{p} = \left \| \mathbf{A}^T \mathbf{A}-\mathbf{I} \right \|_F^2,
\end{equation}
where $\mathbf{A}=\left [ \pmb{\alpha}_{1}, \cdots , \pmb{\alpha}_{N} \right ]$ is the attention matrix; $\mathbf{I}$ is the identity matrix and $\left \| \cdot \right \|_F$ represents the Frobenius norm of a matrix. $\mathcal L_{p}$ can encourage each attention head to extract different information from the same utterance.  It is similar to $L_2$ regularization and is minimized together with the original cost of the system.

\subsection{Loss function}

After getting the utterance-level speaker embedding, we calculate the similarity loss using the generalized end-to-end (GE2E) loss formulation \cite{wan2018generalized}. The GE2E loss is based on processing a large number of utterances at once to minimize the distance of the same speaker while maximizing the distance of different speakers. 

For each batch training, we randomly choose $Q \times P$ utterances from $Q$ different speakers with $P$ utterances per speaker. And we calculate the utterance-level speaker embedding $\tilde{\pmb{e}}_{ji}$ based on Equations \ref{equ_eji}, \ref{equ_att_ht}, \ref{equ_e} for each utterance. $\tilde{\pmb{e}}_{ji}$ represents the speaker embedding of the $j^{\text{th}}$ speaker's $i^{\text{th}}$ utterance. And the centroid of embedding vectors from the $j^{\text{th}}$ speaker is defined:
\begin{equation}
\label{equ_cj}
\mathbf{c}_j =  \mathbf{E}_{i} \left [ \tilde{\pmb{e}}_{ji} \right ] = \frac{1}{P} \sum_{i=1}^{P} \tilde{\pmb{e}}_{ji}.
\end{equation}
GE2E builds a similarity matrix $\mathbf{S}_{ji,k}$ that defines the scaled cosine similarities between each embedding vector $\tilde{\pmb{e}}_{ji}$ to all centroids $\mathbf{c}_k$ $(1 \leqslant j,k \leqslant Q \text{ and } 1 \leqslant i \leqslant P)$:
\begin{equation}
\label{equ_sjik}
\mathbf{S}_{ji,k} = w \cdot \cos(\tilde{\pmb{e}}_{ji}, \mathbf{c}_k) + b,
\end{equation}
where $w$ and $b$ are learnable parameters. The weight is constrained to be positive $w > 0$, because the scaled similarity is expected to be larger when the cosine similarity is larger.

During the training, each utterance's embedding is expected to be similar to the centroid of that utterance's speaker, while far from other speakers' centroids. The loss on each speaker embedding $\tilde{\pmb{e}}_{ji}$ could be defined as:
\begin{equation}
\label{equ_loss_e}
\mathcal L(\tilde{\pmb{e}}_{ji}) = \log \sum_{k=1}^{Q} \exp(\mathbf{S}_{ji,k}) - \mathbf{S}_{ji,j} .
\end{equation}
And the utterance-level GE2E loss $\mathcal L_{u}$ is the sum of all losses over the similarity matrix, shown as:
\begin{equation}
\label{equ_lg}
\mathcal L_{u}(\mathbf{x};\mathbf{w}) = \sum_{j,i}\mathcal L(\tilde{\pmb{e}}_{ji}).
\end{equation}

 For the text-independent speaker verification, each extracted segment-level embedding is expected to capture the speaker characteristics. In order to guide the segment attention to focus on the segments with more speaker discriminations, we further incorporate the similarity
loss of segment-level embeddings.  The segment-level GE2E loss $\mathcal L_{s}$ is similar to the utterance-level GE2E loss $\mathcal L_{u}$ except that it takes all segment-level embeddings as input, which could help the proposed model to learn more effective ways of embedding fusion and accelerate model convergence. The objective function can be formulated as:
\begin{equation}
\label{equ_ls}
\mathcal L_{s}(\mathbf{x};\mathbf{w}) = \sum_{j,i} \sum_{n}\mathcal L(\mathbf{e}_{n}).
\end{equation}

Finally, the utterance-level GE2E loss, segment-level GE2E loss and penalty loss are combined together to construct the total loss, shown as:
\begin{equation} \small
\label{equ_l_all}
\mathcal L = \mathcal L_{u} + \lambda_s \mathcal L_{s} + \lambda_p \mathcal L_{p}
\end{equation}
The magnitude of the segment-level GE2E loss and penalty loss is controlled by hyperparameters $\lambda_s$ and $\lambda_p$. With the joint optimization of the segment-level and utterance-level GE2E loss, both local details and global information of utterances are taken into account. Our proposed method can extract the unified speaker embeddings for utterances of variable duration, which unifies the process of training and testing and avoids the mismatch between them.

\section{Experiments}
\label{sec:Experiments}

We report speaker verification performance on Tongdun and VoxCeleb \cite{nagrani2017voxceleb} corpora. The proposed deep segment attentive embedding is compared with the generalized end-to-end loss based embedding as well as the traditional i-vector. We use Equal Error Rate (EER) to quantify the system performance.

\subsection{Data}
\label{ssec:Data}

 \textbf{Tongdun}. The corpus is from the speaker verification competition held by Tongdun technology company \cite{kesci-web}, which consists of more than $120$K utterances from $1,500$ Chinese speakers in training set and $3,000$ trial pairs are provided as test data. Most of the training data are short utterances with average duration of $3.7$s, while utterances in test set are very long and average duration is about $20$s.

\noindent \textbf{VoxCeleb}. The training set consists of more than $140$K utterances of $1,251$ speakers. And $37,720$ trial pairs from $40$ speakers are used as evaluation data for the verification process. The average duration of training and evaluation data is $8.24$s and $8.28$s, respectively.

For each speech utterance, a VAD \cite{Mak2014A, yu2011comparison} is applied to prune out silence regions. 

\subsection{i-vector system}
\label{ssec:i-vector-system}

The i-vector system uses $20$-dimensional MFCCs as front-end features, which are then extended to $60$-dimensional acoustic features with their first and second derivatives. Cepstral mean normalization is applied. An i-vector of $400$ dimensions is then extracted from the acoustic features using a $2048$-mixture UBM and a total variability matrix. PLDA serves as the scoring back-end. Mean subtraction, whitening, and length normalization \cite{Garcia2011Analysis} are applied to the i-vector as preprocessing steps, and the similarity is measured using a PLDA model with a speaker space of $400$ dimensions.

\subsection{Deep speaker embedding system}
\label{ssec:Deep-speaker-embedding-system}

For deep speaker embedding systems, we take the $40$-dimensional filter-banks with $32\text{-ms}$ Hamming window and $16\text{-ms}$ frame shift as the input features, and each dimension of features is normalized to have zero mean and unit variance over the training set. A combination of $3$-layer LSTM and a linear projection layer is used to extract the speaker embeddings. Each LSTM layer contains $512$ nodes, and the linear projection layer is connected to the last LSTM layer, whose output size is $256$. Therefore, we can extract $256$-dimension speaker embeddings according to the outputs of the linear projection layer. The cosine similarity score of the pair of embedding vectors is computed to verify the speaker. According to \cite{wan2018generalized}, the scaling factors $w$ and $b$ in Eq.~\ref{equ_sjik} are initialized to $10$ and $5$, respectively.

We take the LSTM-based speaker embedding system proposed by Wan \cite{wan2018generalized} as the baseline, which is optimized by GE2E loss. Let us denote the baseline system as ``LSTM-GE2E''. ``LSTM-GE2E'' uses the local segments truncated from utterances to learn the utterance-level speaker embedding. The length of segments is randomly chosen within $[80, 120]$, but all segments in a batch is fixed. In the testing phase, each utterance is segmented by a sliding window of $100$ frames with $50\%$ overlap. We extract the embedding of each segment and then average them as the speaker embedding of the utterance. The embedding of each segment is obtained by performing a frame-level attention pooling operator on the outputs of the linear projection layer.

Compared to ``LSTM-GE2E'', the proposed deep segment attentive embedding system uses the whole utterance to learn the utterance-level speaker embedding by the segment attention, which is denoted as ``DSAE-GE2E''. The segment attention is implemented by performing the multi-head attention pooling on the segment-level embeddings. The attention dim $d_a$ is set to $128$ and the attention head number $d_r$ is chosen from $\left [1, 2, 5 \right]$. In addition, ``DSAE-GE2E'' is jointly optimized by the utterance-level and segment-level GE2E losses, as shown in Eq.~\ref{equ_l_all}. The weights $\lambda_s$ and $\lambda_p$ of terms in Eq.~\ref{equ_l_all} are experimentally set to $0.2$ and $0.001$, respectively.

All deep speaker embedding models are trained from a random initialization by an Adam optimizer \cite{kingma2014adam}. The initial learning rate is set to $0.001$ and decayed according to the performance of the validation set. For each batch training, we randomly choose $640$ utterances of $64$ speakers with $10$ utterances per speaker. We mention that the length of segments in a batch is fixed. About $15,000$ batches are used to train the network. In addition, the $L_{2} \text{ norm}$ of gradient is clipped at $3$ to avoid gradient explosion \cite{pascanu2012understanding}.

\subsection{Results}

In the following results, ``LSTM-GE2E'' refers to the deep speaker embedding system trained with GE2E loss. ``DSAE-GE2E-k'' denotes the proposed deep segment attentive embedding system with the multi-head attention layer of $k$ attention heads.

\begin{table}[tb] \small
  \caption{ \small Speaker Verification Results on Tongdun.}
  \label{tab:eer-Tongdun}
  \centering
  \begin{tabular}{p{3.2cm} p{1.5cm}<{\centering}}
  \specialrule{0em}{3pt}{0pt}
  \Xhline{1.5pt}
  \specialrule{0em}{2pt}{2pt}
  \textbf{Embedding} & \textbf{EER (\%)} \\
  \specialrule{0em}{2pt}{2pt}
  \hline

  \specialrule{0em}{2pt}{2pt}
  i-vector/PLDA & 3.0 \\
  \specialrule{0em}{2pt}{2pt}
  \hline

  \specialrule{0em}{2pt}{2pt}
  LSTM-GE2E  & 2.0 \\
  DSAE-GE2E-1  & 1.5 \\
  DSAE-GE2E-2  & 1.3 \\
  DSAE-GE2E-5  & 1.0 \\
  \specialrule{0em}{2pt}{2pt}
  \Xhline{1.5pt}
  \end{tabular}
\end{table}

Table \ref{tab:eer-Tongdun} shows the performance on Tongdun test set. All deep learning based speaker embedding systems outperform the traditional i-vector system, which shows the effectiveness of the deep speaker embeddings. In general, the proposed ``DSAE-GE2E'' consistently and significantly outperform ``LSTM-GE2E''. For the multi-head attention layer, more attention heads achieve greater improvement. ``DSAE-GE2E-1'' is $25\%$ better in EER than ``LSTM-GE2E'' and ``DSAE-GE2E-5'' outperform ``LSTM-GE2E'' by $50\%$. Note that the difference of durations between Tongdun training and testing utterances is very large and our systems can extract the unified utterance-level speaker embeddings for utterances of variable duration, which significantly improve the system performance. Results indicate that our proposed utterance-level speaker embedding is a duration robust representation for speaker verification.

\begin{table}[t] \small
  \caption{ \small Speaker Verification Results on VoxCeleb.}
  \label{tab:eer-VoxCeleb}
  \centering
  \begin{tabular}{p{3.2cm} p{1.5cm}<{\centering}}
  \specialrule{0em}{3pt}{0pt}
  \Xhline{1.5pt}
  \specialrule{0em}{2pt}{2pt}
  \textbf{Embedding} & \textbf{EER (\%)}  \\
  \specialrule{0em}{2pt}{2pt}
  \hline

  \specialrule{0em}{2pt}{2pt}
  i-vector/PLDA &8.9 \\
  \specialrule{0em}{2pt}{2pt}
  \hline

  \specialrule{0em}{2pt}{2pt}
  LSTM-GE2E  &6.2 \\
  DSAE-GE2E-1  &5.8 \\
  DSAE-GE2E-2  &5.5 \\
  DSAE-GE2E-5  &5.2 \\
  \specialrule{0em}{2pt}{2pt}
  \Xhline{1.5pt}
  \end{tabular}
\end{table}

The performance on VoxCeleb test set is shown in Table \ref{tab:eer-VoxCeleb}. Our proposed ``DSAE-GE2E'' also outperforms the i-vector system and ``LSTM-GE2E'', which demonstrates the effectiveness of the proposed method. ``DSAE-GE2E-1'' is $6.5\%$ better in EER than ``LSTM-GE2E'' and ``DSAE-GE2E-5'' outperform ``LSTM-GE2E'' by $16.1\%$. The relative EER reduction is smaller than Tongdun corpus because there is little duration difference between VoxCeleb training and testing utterances. Our proposed method can obtain greater performance improvement when the difference of durations between training and testing utterances is larger.

\section{Conclusions}
\label{sec:conclusions}

In this paper, we propose the deep segment attentive embedding method to learn the unified speaker embeddings for utterances of variable duration. Each utterance is segmented by a sliding window and LSTM is used to extract the  embedding of each segment. Instead of only using one local segment, we use the whole utterance to learn the utterance-level embedding by applying an attentive pooling to embeddings of all segments. Moreover, the similarity loss of segment-level embeddings is introduced to guide the segment attention to focus on the segments with more speaker discriminations, and jointly optimized with the similarity loss of utterance-level embeddings. Systematic experiments on Tongdun and VoxCeleb demonstrate the effectiveness of the proposed method. In the future work, we will investigate different neural network architectures and attention strategies in order to obtain greater performance improvement.

\section{Acknowledgements}
This work was supported by the China National Nature Science Foundation (No. 61573357, No. 61503382, No. 61403370, No. 61273267, No. 91120303).

{
\footnotesize
\bibliographystyle{IEEEbib}
\bibliography{strings,refs}

\begin{thebibliography}{10}

\bibitem{Dehak2011Front}
Najim Dehak, Patrick~J. Kenny, Réda Dehak, Pierre Dumouchel, and Pierre
  Ouellet,
\newblock ``Front-end factor analysis for speaker verification,''
\newblock {\em IEEE Transactions on Audio Speech and Language Processing}, vol.
  19, no. 4, pp. 788--798, 2011.

\bibitem{Prince2007Probabilistic}
Simon J.~D. Prince and James~H. Elder,
\newblock ``Probabilistic linear discriminant analysis for inferences about
  identity,''
\newblock in {\em IEEE International Conference on Computer Vision}, 2007, pp.
  1--8.

\bibitem{Cumani2013Probabilistic}
Sandro Cumani, Oldřich Plchot, and Pietro Laface,
\newblock ``Probabilistic linear discriminant analysis of i-vector posterior
  distributions,''
\newblock in {\em IEEE International Conference on Acoustics, Speech and Signal
  Processing}, 2013, pp. 7644--7648.

\bibitem{li2017deep}
Chao Li, Xiaokong Ma, Bing Jiang, Xiangang Li, Xuewei Zhang, Xiao Liu, Ying
  Cao, Ajay Kannan, and Zhenyao Zhu,
\newblock ``Deep speaker: an end-to-end neural speaker embedding system,''
\newblock {\em arXiv preprint arXiv:1705.02304}, 2017.

\bibitem{snyder2017deep}
David Snyder, Daniel Garcia-Romero, Daniel Povey, and Sanjeev Khudanpur,
\newblock ``Deep neural network embeddings for text-independent speaker
  verification,''
\newblock in {\em INTERSPEECH}, 2017, pp. 999--1003.

\bibitem{Variani2014Deep}
Ehsan Variani, Xin Lei, Erik Mcdermott, Ignacio~Lopez Moreno, and Javier
  Gonzalez-Dominguez,
\newblock ``Deep neural networks for small footprint text-dependent speaker
  verification,''
\newblock in {\em IEEE International Conference on Acoustics, Speech and Signal
  Processing}, 2014, pp. 4052--4056.

\bibitem{wan2018generalized}
Li~Wan, Quan Wang, Alan Papir, and Ignacio~Lopez Moreno,
\newblock ``Generalized end-to-end loss for speaker verification,''
\newblock in {\em 2018 IEEE International Conference on Acoustics, Speech and
  Signal Processing (ICASSP)}. IEEE, 2018, pp. 4879--4883.

\bibitem{zhang2017end}
Chunlei Zhang and Kazuhito Koishida,
\newblock ``End-to-end text-independent speaker verification with triplet loss
  on short utterances,''
\newblock in {\em Proc. of Interspeech}, 2017.

\bibitem{Sainath2015Convolutional}
T.~N Sainath, O~Vinyals, A~Senior, and H~Sak,
\newblock ``Convolutional, long short-term memory, fully connected deep neural
  networks,''
\newblock in {\em IEEE International Conference on Acoustics, Speech and Signal
  Processing}, 2015, pp. 4580--4584.

\bibitem{Heigold2015End}
Georg Heigold, Ignacio Moreno, Samy Bengio, and Noam Shazeer,
\newblock ``End-to-end text-dependent speaker verification,''
\newblock {\em Computer Science}, pp. 5115--5119, 2015.

\bibitem{okabe2018attentive}
Koji Okabe, Takafumi Koshinaka, and Koichi Shinoda,
\newblock ``Attentive statistics pooling for deep speaker embedding,''
\newblock 2018.

\bibitem{zhu2018self}
Yingke Zhu, Tom Ko, David Snyder, Brian Mak, and Daniel Povey,
\newblock ``Self-attentive speaker embeddings for text-independent speaker
  verification,''
\newblock {\em Proc. Interspeech 2018}, pp. 3573--3577, 2018.

\bibitem{Kenny2013PLDA}
Patrick Kenny, Themos Stafylakis, Pierre Ouellet, Md.~Jahangir Alam, and Pierre
  Dumouchel,
\newblock ``Plda for speaker verification with utterances of arbitrary
  duration,''
\newblock in {\em IEEE International Conference on Acoustics, Speech and Signal
  Processing}, 2013, pp. 7649--7653.

\bibitem{Na2018Deep}
Dan Su Zhifeng~Li Na~Li, Deyi~Tuo and Dong Yu,
\newblock ``Deep discriminative embeddings for duration robust speaker
  verification,''
\newblock in {\em INTERSPEECH}, 2018, pp. 2262--2266.

\bibitem{Nam2018Robust}
Jean-Marc~Odobez Nam~Le,
\newblock ``Robust and discriminative speaker embedding via intra-class
  distance variance regularization,''
\newblock in {\em INTERSPEECH}, 2018, pp. 2257--2261.

\bibitem{Sarthak2018Learning}
Atul~Rai Sarthak~Yadav,
\newblock ``Learning discriminative features for speaker identification and
  verification,''
\newblock in {\em INTERSPEECH}, 2018, pp. 2237--2241.

\bibitem{lin2017structured}
Zhouhan Lin, Minwei Feng, Cicero Nogueira~dos Santos, Mo~Yu, Bing Xiang, Bowen
  Zhou, and Yoshua Bengio,
\newblock ``A structured self-attentive sentence embedding,''
\newblock {\em arXiv preprint arXiv:1703.03130}, 2017.

\bibitem{nair2010rectified}
Vinod Nair and Geoffrey~E Hinton,
\newblock ``Rectified linear units improve restricted boltzmann machines,''
\newblock in {\em Proceedings of the 27th international conference on machine
  learning (ICML-10)}, 2010, pp. 807--814.

\bibitem{nagrani2017voxceleb}
Arsha Nagrani, Joon~Son Chung, and Andrew Zisserman,
\newblock ``Voxceleb: a large-scale speaker identification dataset,''
\newblock {\em arXiv preprint arXiv:1706.08612}, 2017.

\bibitem{kesci-web}
``{Tongdun Technology Speaker Verification Competition},''
  \url{https://www.kesci.com/home/competition/5b4eb2cfe87957000f9024a4/}.

\bibitem{Mak2014A}
Man~Wai Mak and Hon~Bill Yu,
\newblock ``A study of voice activity detection techniques for nist speaker
  recognition evaluations,''
\newblock {\em Computer Speech and Language}, vol. 28, no. 1, pp. 295--313,
  2014.

\bibitem{yu2011comparison}
Hon-Bill Yu and Man-Wai Mak,
\newblock ``Comparison of voice activity detectors for interview speech in nist
  speaker recognition evaluation,''
\newblock in {\em Twelfth Annual Conference of the International Speech
  Communication Association}, 2011.

\bibitem{Garcia2011Analysis}
Daniel Garcia-Romero and Carol~Y. Espy-Wilson,
\newblock ``Analysis of i-vector length normalization in speaker recognition
  systems,''
\newblock in {\em INTERSPEECH 2011, Conference of the International Speech
  Communication Association, Florence, Italy, August}, 2011, pp. 249--252.

\bibitem{kingma2014adam}
Diederik~P Kingma and Jimmy Ba,
\newblock ``Adam: A method for stochastic optimization,''
\newblock {\em arXiv preprint arXiv:1412.6980}, 2014.

\bibitem{pascanu2012understanding}
Razvan Pascanu, Tomas Mikolov, and Yoshua Bengio,
\newblock ``Understanding the exploding gradient problem,''
\newblock {\em CoRR, abs/1211.5063}, 2012.

\end{thebibliography}
}

\end{document}